\documentclass[
	aps,
	prb,
	twocolumn, 		% for Phys. Rev. appearance, change preprint to twocolumn
	superscriptaddress, 	% use superscriptaddress for long author lists, or if there are many overlapping affiliations
	showpacs,		% add 'showpacs' option to make PACS codes appear
	showkeys,		        % add 'showkeys' option to make keywords appear
	secnumarabic,
	amsmath,
	amssymb,
	floatfix,
	a4paper,
	balancelastpage,
]{revtex4-1}

\usepackage[latin1]{inputenc}% ISO Latin 1 encoding (allows the use of non-ascii characters such as in "Mössbauer" or Jülich
\usepackage{amssymb}
\usepackage{graphicx}
\usepackage{dcolumn}		% Align table columns on decimal point
\usepackage{bm}			% bold math
\usepackage{setspace}  		% file:///usr/local/texlive/2012/texmf-dist/tex/latex/setspace/setspace.sty 
\usepackage{units}
\usepackage{color,soul}  		% use \hl{text here} to highlight text
\newcommand {\Fig} {Fig.~}      % do not highlight "Fig."
\newcommand {\dg} {\ensuremath{^{\circ}}}
\newcommand {\mub} {\ensuremath{\mu_{\text{B}}}}
\newcommand {\xray} {\hbox{x-ray}}
\newcommand {\etal} {\textit{et al.}}
\newcommand {\TK} {\ensuremath{T_{K}}}
\newcommand {\TN} {\ensuremath{T_{N}}}

\newcommand {\ECG} {EuCu$_{2}$Ge$_{2}$}
\newcommand {\ECS} {EuCu$_{2}$Si$_{2}$}
\newcommand {\ECSGx} {EuCu$_{2}$(Si$_{x}$Ge$_{1-x}$)$_{2}$}

\begin{document}

%\title{Novel aspects of the coexistence between long-range magnetic order and spin fluctuations in rare-earth intermetallics: a study of \ECSGx}
\title{Europium mixed-valence, long-range magnetic order, and dynamic magnetic response in \ECSGx}

\author{K.~S.~Nemkovski} 
\email{k.nemkovskiy@fz-juelich.de} 
\affiliation{ Jülich Centre for Neutron Science JCNS at Heinz Maier-Leibnitz Zentrum MLZ, Forschungszentrum Jülich GmbH, Lichtenbergstra{\ss}e 1, 85747 Garching, Germany}

\author{D.~P.~Kozlenko}
\affiliation{Frank Laboratory of Neutron Physics, 141980 JINR, Joliot-Curie 6, Dubna, Moscow Region, Russia}

\author{P.~A.~Alekseev} 
\affiliation{National Research Centre ``Kurchatov Institute'', Kurchatov sqr. 1, 123182 Moscow, Russia}
\affiliation{National Research Nuclear University MEPhI (Moscow Engineering Physics Institute), Kashirskoe shosse 31, 115409, Moscow, Russia}

\author{J.-M.~Mignot} 
\affiliation{Laboratoire L\'eon Brillouin - UMR12 CNRS-CEA, CEA Saclay, 91191 Gif-sur-Yvette, France}

\author{A.~P.~Menushenkov} 
\affiliation{National Research Nuclear University MEPhI (Moscow Engineering Physics Institute), Kashirskoe shosse 31, 115409, Moscow, Russia}

\author{A.~A.~Yaroslavtsev} 
\affiliation{National Research Nuclear University MEPhI (Moscow Engineering Physics Institute), Kashirskoe shosse 31, 115409, Moscow, Russia}
\affiliation{European XFEL GmbH, Holzkoppel 4, 22869 Schenefeld, Germany}

\author{E.~S.~Clementyev} 
\affiliation{REC ``Functional Nanomaterial'', I. Kant Baltic Federal University, Nevskogo Str., 14A , 236041 Kaliningrad, Russia}
\affiliation{Institute for Nuclear Research RAS, 117312 Moscow, Russia}

\author{A.~S.~Ivanov} 
\affiliation{Institut Laue-Langevin, BP 156, 38042 Grenoble Cedex 9, France}

\author{S.~Rols} 
\affiliation{Institut Laue-Langevin, BP 156, 38042 Grenoble Cedex 9, France}

\author{B.~Klobes} 
\affiliation{Jülich Centre for Neutron Science JCNS and Peter Grünberg Institute PGI, JARA-FIT, Forschungszentrum Jülich GmbH, 52425 Jülich, Germany}

\author{R.~P.~Hermann} 
\affiliation{Materials Science \& Technology Division, Oak Ridge National Laboratory, P.O. Box 2008, TN 37831-6064, Oak Ridge, USA}

\author{A.~V.~Gribanov} 
\affiliation{Chemistry Department of the Moscow State University, Leninskie Gory, GSP-1, 119991 Moscow, Russia}

\date{\today}

\begin{abstract}
In mixed-valence or heavy-fermion systems, the hybridization between local $f$ orbitals and conduction band states can cause the suppression of long-range magnetic order, which competes with strong spin fluctuations. Ce- and Yb-based systems have been found to exhibit fascinating physical properties (heavy-fermion superconductivity, non-Fermi-liquid states, etc.) when tuned to the vicinity of magnetic quantum critical points by use of various external control parameters (temperature, magnetic field, chemical composition). Recently, similar effects (mixed-valence, Kondo fluctuations, heavy Fermi liquid) have been reported to exist in some Eu-based compounds. Unlike Ce (Yb), Eu has a multiple electron (hole) occupancy of its $4f$ shell, and the magnetic Eu$^{2+}$ state ($4f^7$) has no orbital component in the usual $LS$ coupling scheme, which can lead to a quite different and interesting physics. In the \ECSGx\ series, where the valence can be tuned by varying the Si/Ge ratio, it has been reported that a significant valence fluctuation can exist even in the magnetic order regime. This paper presents a detailed study of the latter material using different microscopic probes (XANES, Mössbauer spectroscopy, elastic and inelastic neutron scattering), in which the composition dependence of the magnetic order and dynamics across the series is traced back to the change in the Eu valence state. In particular, the results support the persistence of valence fluctuations into the antiferromagnetic state over a sizable composition range below the critical Si concentration $x_c \approx 0.65$. The sequence of magnetic ground states in the series is shown to reflect the evolution of the magnetic spectral response.
\end{abstract}


\maketitle

\section{\label{sec:intro}Introduction}
Rare-earth intermetallic compounds provide unique opportunities for furthering our understanding of magnetism in solids. Systems containing rare-earth elements  (Ce, Sm, Eu, Tm, Yb) with unstable $4f$-shells exhibit challenging physical phenomena, such as Kondo effect, electron mass enhancement, valence fluctuations, unconventional (magnetically driven) superconductivity, non-Fermi liquid state, or critical fluctuations in the vicinity of a quantum critical point.\cite{Lawrence'81,Riseborough'00,Aynajian'12,Mathur'98,Coleman'07,Stockert'11,Jang'14,Moll'15,Dahm'09,Yu'09} The possibility for lanthanide-ion magnetism to depart from the canonical Russel-Saunders$+$spin-orbit$+$crystal-field ionic coupling scheme has been extensively documented since the discovery of so-called ``unstable-valence" materials (SmB$_{6}$, SmS) back in the 1960s.\cite{Vainshtein'65,Paderno'67,Menth'69,Nickerson'71,Bucher'71,Kaldis'72,Tao'75} Considerable experimental and theoretical effort has been devoted to those materials over the last decades. Our current understanding of their properties is based, to a large extent, on the so-called ``periodic Anderson model'', consisting of a narrow band of localized $4f$ electrons subject to strong Coulomb correlations, hybridized with a broad conduction band of itinerant $sd$-electrons. Depending on the hybridization strength, band structure, or electronic configuration, a large variety of situations is predicted, which may account for some of the aforementioned experimental properties. Non-Fermi liquid behaviors usually occur in the region of parameter space where magnetic long-range order (LRO) becomes destabilized by spin fluctuations or alternative ``local quantum critical" phenomena.\cite{Dahm'09,Yu'09,Schroder'00,Gegenwart'08}

Most studies in that field have focused on Ce- and Yb-based heavy-fermion (HF) compounds. In particular the $RT_{2}X_{2}$ type ($T$: $3d$ or $4d$ transition metal, $X$: Si, Ge), and CeCu$_{6-x}$Au$_{x}$ families, provide numerous examples of the interplay between long-range magnetic order and Kondo fluctuations considered in Doniach's seminal work on Kondo lattices,\cite{Doniach'77} leading to the discovery of novel quantum critical phenomena. In such systems, nearly trivalent lanthanide ions have single-electron (or -hole) occupancy of their $4f$ shells. As the hybridization becomes stronger, and the $4f$ level approaches the Fermi energy, charge instability sets in and the average rare-earth valence significantly deviates from $+3$. The electronic state can then be described as a quantum superposition of the 4$f^{n}$ and 4$f^{n-1}$ + [5$d$-6$s$]$^{1}$ configurations.\cite{Lawrence'81} In that regime, long-range magnetic order at low temperature is normally hindered by short-lived fluctuations, and a strongly damped dynamical magnetic response results as seen, e.g., in CePd$_{3}$ (Ref.~\onlinecite{HollandMoritz'77}) or YbAl$_{3}$.\cite{Murani'85}

Valence instability has also been found to occur in rare-earth elements with multiple 4$f$-shell occupancies, such as Sm, Eu, Tm, and possibly Pr. A limited number of examples are known, among which archetypal Kondo insulators, such as SmB$_{6}$ or YbB$_{12}$. Recently, the existence of a HF state in Eu-based compounds has attracted renewed interest.\cite{Hossain'04,Danzenbacher'09,Hiranaka'13,Mitsuda'13,Guritanu'12,Hotta'15} This behavior has been clearly evidenced, in particular for the MV compound EuNi$_2$P$_2$ ($v = 2.45$--2.55), from  thermodynamic and transport,\cite{Hiranaka'13} as well as spectroscopic measurements.\cite{Danzenbacher'09,Guritanu'12} Hossain \etal \cite{Hossain'04} have also reported the observation of Kondo effect with a HF behavior in the \ECSGx\ series. Those experimental results raise very interesting questions regarding the applicability to Eu systems of interpretations originally devised for Ce or Yb. The materials studied belong to the same structural class of so-called ``1-2-2'' rare-earth intermetallics as the above-mentioned $RT_{2}X_{2}$ systems ($R$: Ce, Yb). Eu-based 1-2-2 compounds have actually been found to exhibit a variety of unconventional behaviors: hybridization gap formation in EuNi$_{2}$P$_{2}$,\cite{Danzenbacher'09} reentrant superconductivity under pressure, competing with long-range Eu magnetic order, in EuFe$_{2}$As$_{2}$,\cite{Kurita'11} along with a valence instability of Eu.\cite{Sun'10}

In the \ECSGx\ series, Eu occurs in a composition-dependent mixed-valence (MV) state. The ground-state multiplets of the two parent ionic configurations are $^{7}F_{0}$ (nonmagnetic, $J = 0$) for Eu$^{3+}$, and $^{8}S_{7/2}$ (spin-only, $J = 7/2$) for Eu$^{2+}$. In pure \ECG\ and the solid solutions with $0 \leq x \leq 0.6$, transport and thermodynamics measurements \cite{Hossain'04} evidence a phase transition around 15 K, which is ascribed to magnetic ordering with a magnetic moment estimated \cite{Alekseev'14} to approach the theoretical value for Eu$^{2+}$. Pure \ECS, on the other hand, can be described as an intermediate-valence Van-Vleck paramagnet. The suppression of magnetic order and the transition to a Fermi-liquid, HF regime takes place in a concentration range, near x$_c \sim 0.65$, where the Eu valence was reported\cite{Hossain'04,Fukuda'03} to deviates strongly from an integer value. In the Si-rich compounds ($x =0.9, 1.0$), inelastic neutron scattering (INS) spectra\cite{Alekseev'12} are characterized by a renormalized Eu$^{3+}$-like intermultiplet (spin-orbit) transition, together with an extra magnetic peak at lower energy, which has been interpreted as an exciton-like ``resonance'', related to the formation of a spin gap of 20--30 meV below $T \sim 100$ K. 

In this work, we address the question of how the competing Kondo and magnetic ordering phenomena reported in Ref.~
\onlinecite{Hossain'04} compare to those studied previously in Ce or Yb compounds. The key issue of a possible coexistence of long-range magnetic order with a MV state, as suggested in previous work,\cite{Hossain'04,Fukuda'03} is addressed by means of different microscopic probes (XANES, Mössbauer spectroscopy, neutron powder diffraction (NPD)). Evidence is reported for an unconventional coexistence of long-range magnetic order and a homogeneous MV state with spin fluctuations in \ECSGx, occurring over a significant range of Si concentrations $x$ below the critical value $x_c \approx 0.65$. This behavior is at variance with the general trend observed in other unstable-valence compounds.

 The evolution of the magnetic spectral response across the \ECSGx\ series was studied in a wide temperature range using INS experiments. In particular, high-resolution time-of-flight measurements on the Ge-rich compounds reveal the existence of narrow, concentration-dependent, quasielastic (QE) signal. On decreasing the Si concentration, the gradual change in the magnetic relaxation rate, indicated by the narrowing of this spectral component, together with the strong renormalization to lower energies of the Eu$^{3+}$ spectral contribution, is found to play a key role in the formation of the unconventional magnetic and HF states in the vicinity of $x_c$.

\section{\label{sec:exp}Experimental details}

\subsection{\label{ssec:sprep} Sample preparation}
The \ECSGx\ samples used in this work were prepared by arc melting from high-purity materials, Si, Ge, Cu ($>$ 99.99\%) and Eu (99.9\%). All of them were annealed at $T_{ann} = 0.8 T_{melt}$ (melting temperature) during $\sim 200$ hours. Further characterization by \xray\ powder diffraction showed that all samples crystallized in the body-centered tetragonal ThCr$_{2}$Si$_{2}$-type structure ($I4/mmm$ space group, \#139)
No impurity phase was detected within the sensitivity of the method.

\subsection{\label{ssec:xanes} X-ray absorption and Mössbauer spectroscopy}
The valence state of the \ECSGx\ compounds ($0 \leq x \leq 0.9$) was determined by \xray\ absorption near-edge structure (XANES) spectroscopy at the Eu $L_{3}$ edge, and by Mössbauer spectroscopy. These methods are known to probe inter-configurational valence fluctuations on very different time scales: about $10^{-8}$ s for Mössbauer spectroscopy, as compared to $\sim 10^{-15}$ s for XANES. The XANES measurements were performed on the A1 beamline of the DORIS-III storage ring (DESY Photon Science, Hamburg) and at the $mySpot$ beamline of BESSY-II (HZB, Berlin) in transmission geometry. In the data treatment, the $f^{6}$ and $f^{7}$ components were described by Lorentzian profiles at their respective centre positions, whereas photoelectron excitations to the continuous spectrum could be represented by an arctangent function. To simulate experimental broadening, the overall function was convoluted with a Gaussian distribution. The experimental temperature range was 7 K $\leq T \leq 300$ K.

$^{151}$Eu Mössbauer spectra were collected at the Forschungszentrum Jülich on a constant-acceleration spectrometer using a 30 mCi $^{151}$SmF$_{3}$ source. The velocity calibration was performed with $\alpha$-Fe at room temperature (RT), using a $^{57}$Co/Rh source. All Mössbauer spectra discussed here were obtained at RT on powder samples, and the isomer shifts (IS) are derived with reference to EuF$_{3}$.

\subsection{\label{ssec:npd} Neutron powder diffraction}
Three \ECSGx\ powder samples with compositions $x = 0.0$, 0.40, and 0.60, corresponding to the part of magnetic phase diagram where magnetic order is expected to occur, were measured on the hot-neutron diffractometer 7C2 at LLB-Orphée in Saclay. The sample masses were 0.66, 0.52, and 0.69 g for the three above compositions, respectively. Neutron scattering experiments on compounds containing natural Eu are challenging because of the very large absorption cross section of Eu ($\sigma_{abs} = 4530$ b for 2200 m/s neutrons). However, this problem can be circumvented by using incoming neutrons with a relatively short wavelength, $\lambda = 1.121$ \AA, from a Ge(111) monochromator. The samples were prepared in a slab geometry, with an area of $12\times45$ mm$^{2}$ and a thickness of approximately 0.3 mm, corresponding to 0.15 mm of bulk material. Sample powder was packed in flat-shaped thin-foil Al sachets, whose surface was oriented perpendicular to the incoming monochromatic neutron beam. The scattering angle range in which intense peaks are observed was $3.3\dg < 2\Theta < 40\dg$, which corresponds to a momentum transfer range $0.3 < Q < 3.8$ \AA$^{-1}$. Under those conditions, a nearly constant level of the transmission could be achieved, varying from 75\% to 72\% in the entire scattering angle range of interest. For all samples, diffraction patterns have been recorded at temperatures comprised between 4 K and 50 K using an ILL-type Orange cryostat. The data analysis was performed using the Rietveld refinement program \textsc{FullProf}.\cite{fullprof'93,fullprof'01} 

A fourth sample, with $x = 0.75$, was not measured on 7C2 for lack of experimental beam time, but the absence of magnetic Bragg peaks for this composition was deduced from the analysis of the elastic signal in the time-of-flight measurements on IN4C (see Section \ref{sec:results}).

\subsection{\label{ssec:nspec} Neutron spectroscopy}
INS experiments were carried out on the thermal-neutron time-of-flight spectrometer IN4C at the ILL in Grenoble, with a resolution of 1.65 meV (FWHM at zero energy transfer, from the width of the vanadium elastic line). The measurements were performed using incident neutrons at energy $E_i = 36.3$ meV ($\lambda = 1.5$ \AA), from a PG(004) monochromator. For that energy, a transmission factor of about 50\% was achieved using thin samples ($\sim 0.3$ mm of powder). With about 0.8 g of material in the beam, the typical measuring time for one spectrum was about 10 hours.

\section{\label{sec:results}Results}

\subsection{\label{ssec:resxanes} XANES and Mössbauer spectra}
\begin{figure*} %---------------------------------------------------------- Figure XANES and Mössbauer ---------------------------------------------------------------------------
%\vspace{-0.6cm}
\centering
\includegraphics[angle=0,width=0.33\textwidth]{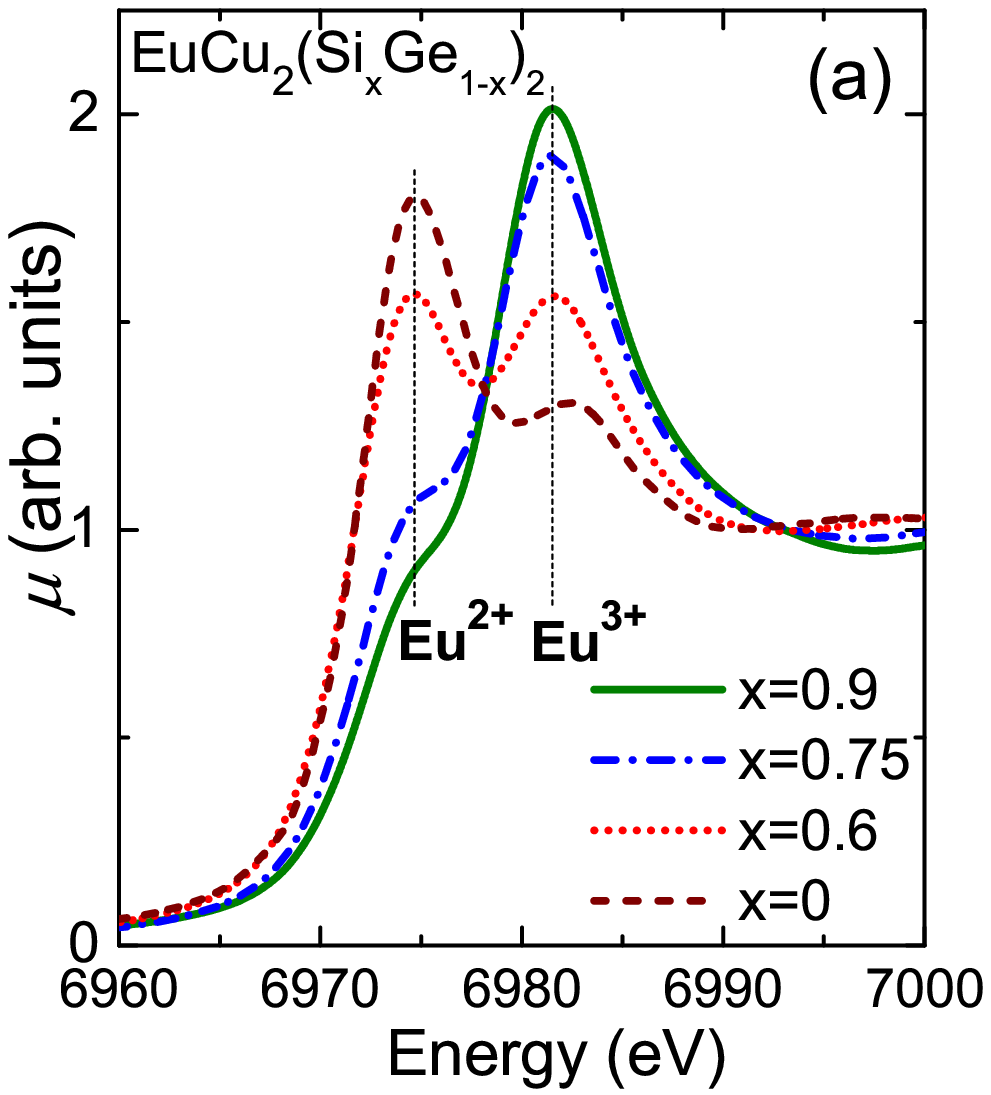}
\includegraphics[angle=0,width=0.36\textwidth]{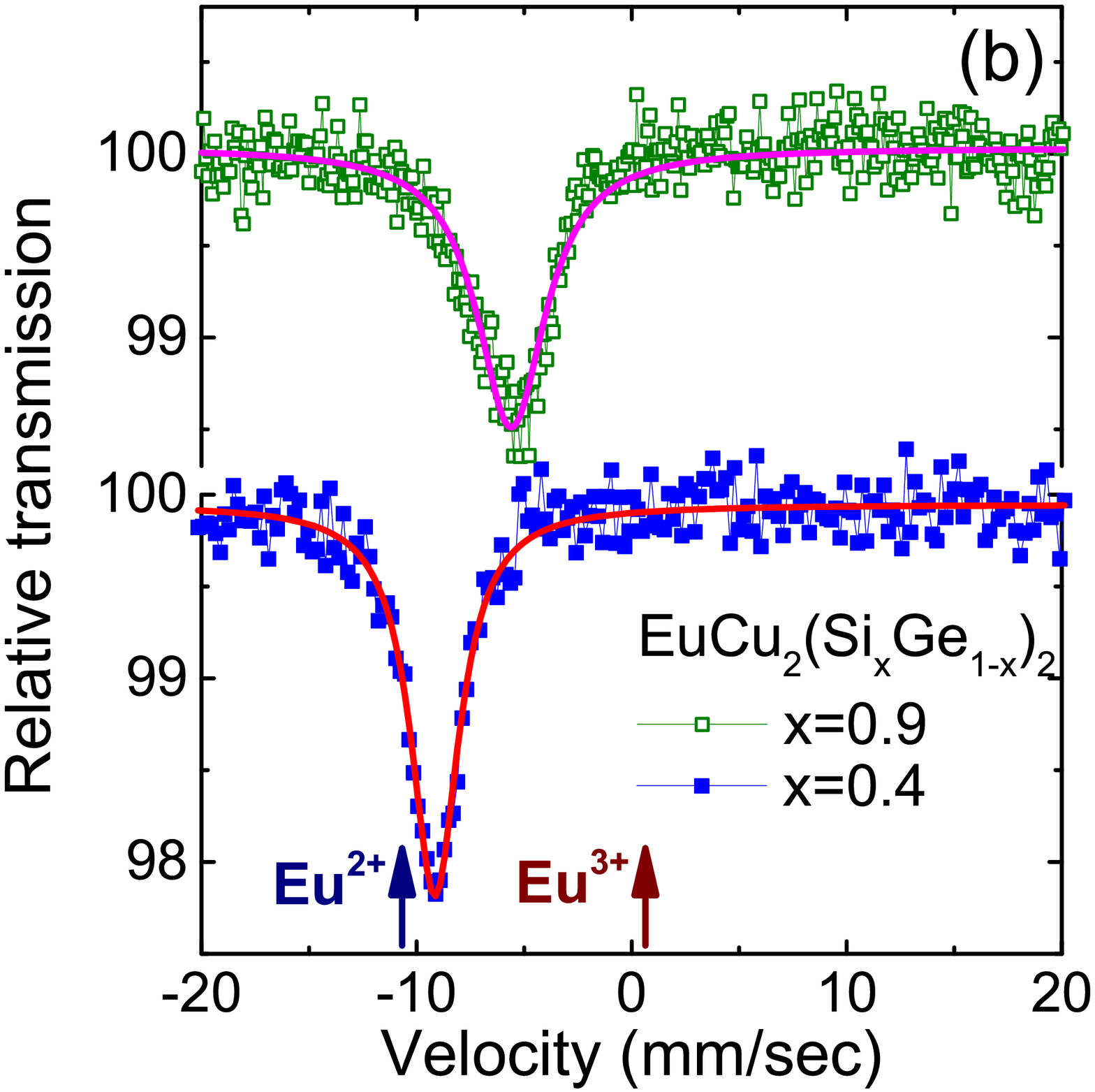}
\includegraphics[angle=0,width=0.40\textwidth]{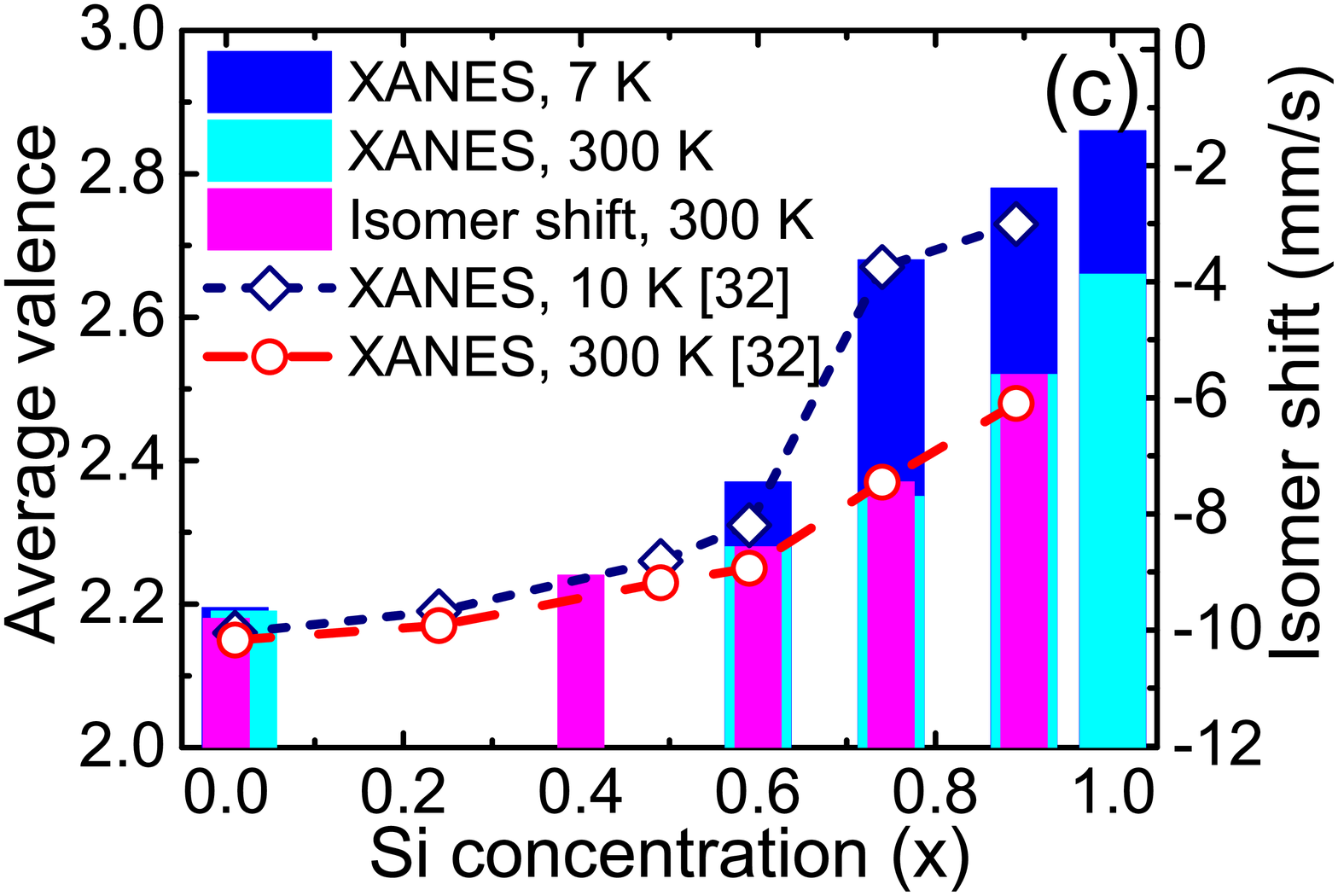}
\caption{\label{xan-moss} (Color online) (a) XANES transmission spectra in \ECSGx\ for $x = 0$, 0.6, 0.75, and 0.9 at $T = 7$ K. 
(b) $^{151}$Eu Mössbauer absorption spectra for $x = 0.6$ and 0.9 at RT, fitted to a single Lorentzian profile; arrows indicate the positions expected for the pure Eu$^{2 +}$ and Eu$^{3 +}$ electronic configurations, derived from measurements on the isostructural compounds Eu$^{2+}$Pd$_{2}$Ge$_{2}$ ($-10.6$ mm/s) and Eu$^{3+}$Ru$_{2}$Si$_{2}$ ($+0.6$ mm/s).\cite{Sampathkumaran'81} (c) Bar chart showing the average Eu valence as a function of the Si concentration derived from the present XANES data for $T = 7$ K (blue) and 300 K (cyan), as well as from the Mössbauer IS at $T = 300$ K (pink).The dashed lines with symbols show the values determined by Fukuda \etal \cite{Fukuda'03} from XANES at $T = 10$\,K (short-dashed dark-blue line with diamonds) and 300 K (red dashed line with circles).}
\end{figure*}

XANES measurements at the Eu $L_{3}$ edge between 7 K and 300 K have been performed on \ECSGx\ for $x = 0$, 0.6, 0.75, 0.9, and 1.0. The spectra for the three solid solutions are shown in \Fig \ref{xan-moss}(a). In the Si-rich samples, the main peak from the Eu$^{3+}$ electronic configuration exhibits a shoulder at lower energy indicating a sizable Eu$^{2+}$ contribution. With decreasing Si concentration, the contribution of this Eu$^{2+}$ component gradually increases to the expense of the Eu$^{3+}$ component, and becomes dominant for $x < 0.6$. The Eu valence for each composition was derived from the relative spectral weights by fitting the spectra as explained in Section~\ref{ssec:xanes}). This procedure is based on the assumption that final-state (``shake-up'') effects can be neglected, which has been questioned in previous studies of Eu intermetallics.\cite{Sampathkumaran'85,Michels'94} This point is discussed in more detail hereafter (Section~\ref{sec:discuss}). The obtained composition dependence of the Eu valence at $T = 7$ K and 300 K, plotted in \Fig \ref{xan-moss}(c), is quite consistent with the previous data of Fukuda \etal. \cite{Fukuda'03}

%---------------------------------------------------------------------------------------------------------------------------------------------------------------------------------------------------

The $^{151}$Eu Mössbauer spectra at RT are characterized, for all studied concentrations ($x = 0$, 0.4, 0.6, 0.75, and 0.9) by a single absorption line, as illustrated in \Fig \ref{xan-moss}(b) for $x = 0.4$ and 0.9. The position of this line (Mössbauer IS) is clearly intermediate between those expected for pure Eu$^{2+}$ and Eu$^{3+}$ electronic configurations (denoted by arrows in the plot). Its lineshape is well described by a single Lorentzian function, which implies that the Eu$^{2+}$--Eu$^{3+}$ valence mixing state is \textit{homogeneous} on the characteristic time scale of the measurement ($\sim$ 10$^{-8}$ s).

From the composition dependence of the IS, it is clear that the average valence at RT steadily increases with increasing Si content. To get a quantitative estimate, one needs to know precisely the positions expected for pure Eu$^{2+}$ and Eu$^{3+}$ valence states, which actually depend on the unit cell volume, and therefore vary slightly from one family of compounds to another.\cite{Wortmann'85,Nowik'87} Here we limited ourselves to checking the consistency of the composition dependence of the Eu valence derived from the IS, by normalizing the values obtained for $x = 0.6$ and 0.9  to those derived from XANES (2.28 and 2.52). The resulting agreement between the two methods in the entire composition range, shown in \Fig \ref{xan-moss}(c), is satisfactory.

\subsection{\label{ssec:lrmo} Long-range magnetic order}
\begin{figure}[t] %---------------------------------------------------------- Figure neutron diffraction ---------------------------------------------------------------------------------
%\centering
\includegraphics[angle=0,width=0.90\columnwidth]{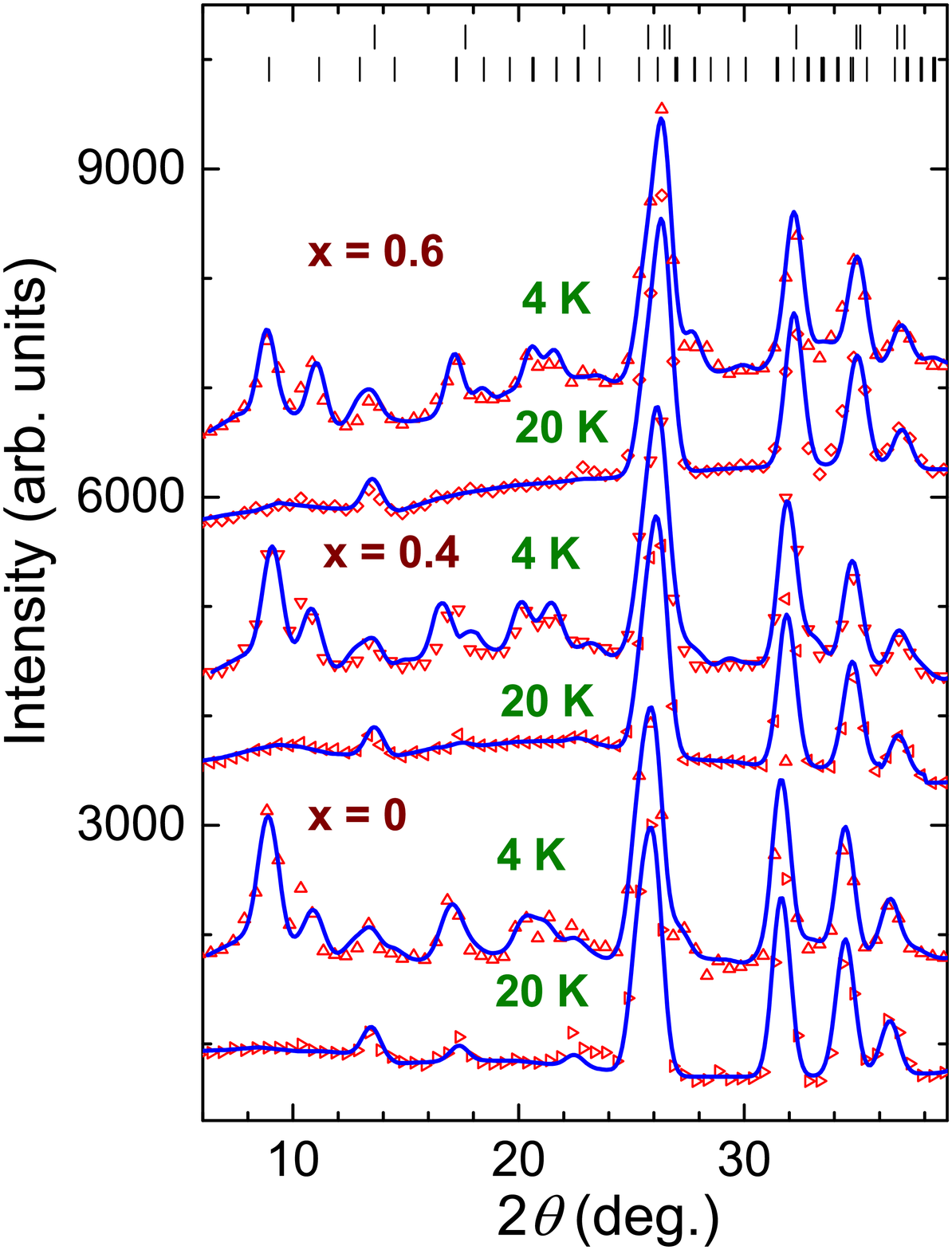}
\caption{\label{diffpat} (Color online) Neutron diffraction patterns of \ECSGx\ ($x = 0$, 0.4, 0.6) measured at $T = 20$ K  and 4 K and refined using the Rietveld method. The ticks at the top of the frame represent the calculated positions of the nuclear reflections corresponding to the tetragonal $I4/mmm$ crystal structure (upper row) and the magnetic reflections corresponding to the propagation vector $\bm{k} = (\frac13,\,0,\,0)$ (lower row).} 
\end{figure}

For the three measured concentrations $x = 0$, 0.4, 0.6, the diffraction patterns collected at $T_{min} \approx 4$ K show clear evidence of magnetic superstructure reflections, which vanish in the paramagnetic phase (\Fig \ref{diffpat}). These satellites demonstrate that long-range magnetic order occurs between \ECG\ and EuCu$_{2}$Si$_{1.2}$Ge$_{0.8}$. For $x = 0.75$, on the other hand, careful analysis of the elastic signal in our time-of-flight measurements on IN4C (Section~\ref{ssec:nspec}) revealed no Bragg satellites indicative of magnetic order, whereas such satellites are clearly seen in EuCu$_{2}$Si$_{1.2}$Ge$_{0.8}$ under the same experimental conditions. The present results thus agree perfectly with those obtained previously from thermodynamic and transport measurements,\cite{Hossain'04} where the critical concentration of the long range magnetic order suppression was estimated to be $x_{c} = 0.65$.

%---------------------------------------------------------------------------------------------------------------------------------------------------------------------------------------------------

For the three compositions showing magnetic order, the superstructure peaks can be indexed using a single, commensurate, magnetic wave vector $\bm{k} = (\frac13,\,0,\,0)$. The data refinement points to the formation of a spin spiral antiferromagnetic (AFM) structure, in which the Eu magnetic moments located at the $2a$ Wyckoff positions, $(0,\,0,\,0)$ and $(\frac12,\,\frac12,\,\frac12)$, in the tetragonal unit cell of the $I$4/$mmm$ space group, are antiparallel. The refined ordered magnetic moment is smaller in the two diluted systems (5.3 \mub) than in pure \ECG\ (6.7  \mub). In a recent NPD study of undoped \ECG, Rowan-Weetaluktuk \etal \cite{RowanW'14} have reported a magnetically inhomogeneous ground state consisting of two incommensurate AFM phases. The reason for this discrepancy is not known but it might be suggested that even a small amount (below the limit of detection of NPD) of strongly dispersed EuO impurity phase, with a very large Eu$^{2+}$ magnetic moment, could significantly affect the ordered magnetic state.

\begin{figure} %------------------------------------------------------- Figure INS background correction ---------------------------------------------------------------------------
%\vspace{-0.6cm}
%\centering
\includegraphics[angle=0,width=0.90\columnwidth]{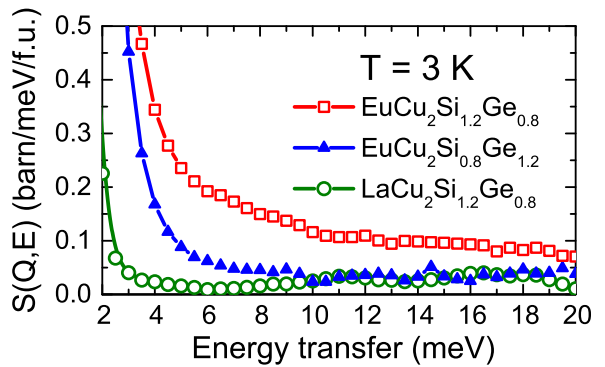}
%\vspace{-0.1cm}
\caption{\label{ins-bkgdcorrec} Time-of-flight INS spectra of \ECSGx\ at $T = 3$ K, measured on IN4C with incident neutron energy $E_{0} = 36.3$ meV (resolution at zero energy transfer $\Gamma = 1.65$ meV, FWHM). Intensities have been averaged over the scattering angle range 13\dg -- 32\dg.  Data for $x = 0.6$ (red squares) and 0.4 (blue triangles) are displayed, together with those (black circles) obtained for LaCu$_2$Si$_{1.2}$Ge$_{0.8}$, which serve as an estimate of the nuclear background contribution.}
%\vspace{-0.6cm}
\end{figure}
%---------------------------------------------------------------------------------------------------------------------------------------------------------------------------------------------------

\subsection{\label{ssec:nscat} Neutron scattering spectra}
In this section, we present the results of the time-of-flight INS experiments performed on IN4C. The magnetic contribution to the inelastic scattering was  obtained experimentally as the difference between the data measured on the \ECSGx\ samples and on the nonmagnetic reference compound LaCu$_2$Si$_{1.2}$Ge$_{0.8}$. An example of the spectra for $T = 3$ K is shown in \Fig \ref{ins-bkgdcorrec}. One sees that the magnetic signal can be determined reliably even for the composition EuCu$_{2}$Si$_{0.8}$Ge$_{1.2}$  ($x = 0.4$) at which its intensity is the weakest. 

\begin{figure} %----------------------------------------------------------- Figure INS magnetic spectra ------------------------------------------------------------------------------
%\vspace{-0.6cm}
%\centering
\includegraphics[angle=0,width=0.4\textwidth]{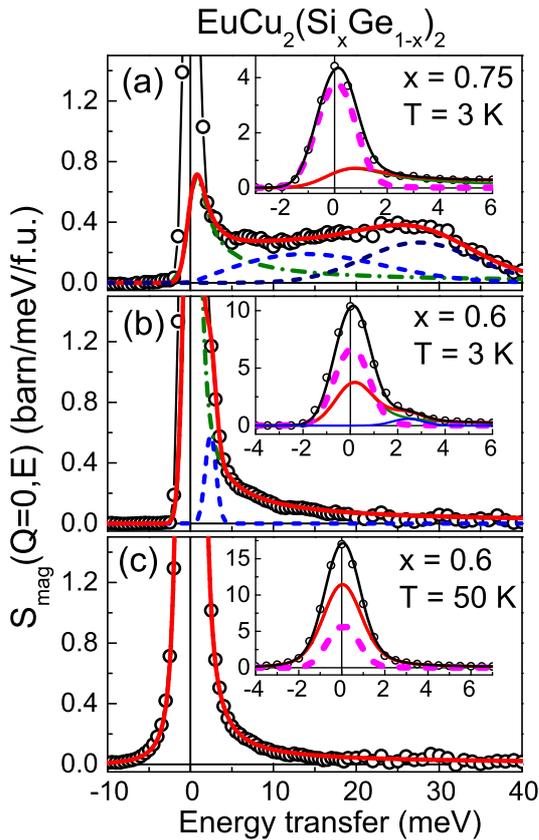}
%\vspace{-0.1cm}
\caption{\label{ins-mag} (Color online) Magnetic neutron scattering spectra of EuCu$_{2}$Si$_{1.5}$Ge$_{0.5}$ ($x = 0.75$) at $T = 3$ K (a), and of EuCu$_{2}$Si$_{1.2}$Ge$_{0.8}$ ($x = 0.6$) at $T = 3$ K (b) and 50 K (c), combined from those measured in present experiment on IN4C at $E_{0} = 36.3$ meV (-10 $\leq$ E $\leq$ 20 meV) and previous study on MARI at ISIS\cite{Alekseev'12} at $E_{0} = 100$ meV (20 $\leq$ E $\leq$ 40 meV). Black circles: experimental values after vanadium normalization, background correction, and subtraction of the nuclear (incoherent elastic and phonon) scattering, estimated from measurements of nonmagnetic LaCu$_{2}$Si$_{1.2}$Ge$_{0.8}$. Lines: (solid red) total magnetic signal; (dashed-dotted green) quasielastic line fitted to a Lorentzian lineshape (temperature factor included), convoluted with a Gaussian resolution function);\footnote{In frame (c), the red and green traces are superimposed.} (dashed blue and dark blue) inelastic response. Insets: same data on a larger intensity scale, further showing the elastic peak with a vanadium lineshape (in magenta, dashed). This residual nuclear scattering signal results from the fact that the large nuclear incoherent scattering cannot be determined with sufficient accuracy from the La reference compound. Nonetheless, the value obtained is consistent between the different samples.}
%\vspace{-0.6cm}
\end{figure}

The magnetic spectra of EuCu$_{2}$Si$_{1.5}$Ge$_{0.5}$ ($x = 0.75$), measured at $T = 3$ K, and of EuCu$_{2}$Si$_{1.2}$Ge$_{0. 8}$ ($x = 0.6$), measured at $T = 3$ and 50 K, are shown in \Fig \ref{ins-mag} in combination with data previously collected on MARI at ISIS\cite{Alekseev'12}. With an incident neutron energy of $E_{0} = 36.3$ meV, the resolution at zero energy transfer was $\Gamma = 1.65$ meV, FWHM), giving access to the low-energy part of the magnetic response, which is the main focus of this study and was not addressed in earlier experiments.\cite{Alekseev'12} All spectra have been reduced to $Q = 0$ according to the magnetic form factor for the $^7F_0 \rightarrow ^7F_1$ spin-orbit transition of  Eu$^{3+}$.

%---------------------------------------------------------------------------------------------------------------------------------------------------------------------------------------------------

\begin{figure*} %------------------------------------------------------------- Figure INS quasielastic --------------------------------------------------------------------------------
%\vspace{-0.6cm}
%\centering
\includegraphics[angle=0,width=0.46\textwidth]{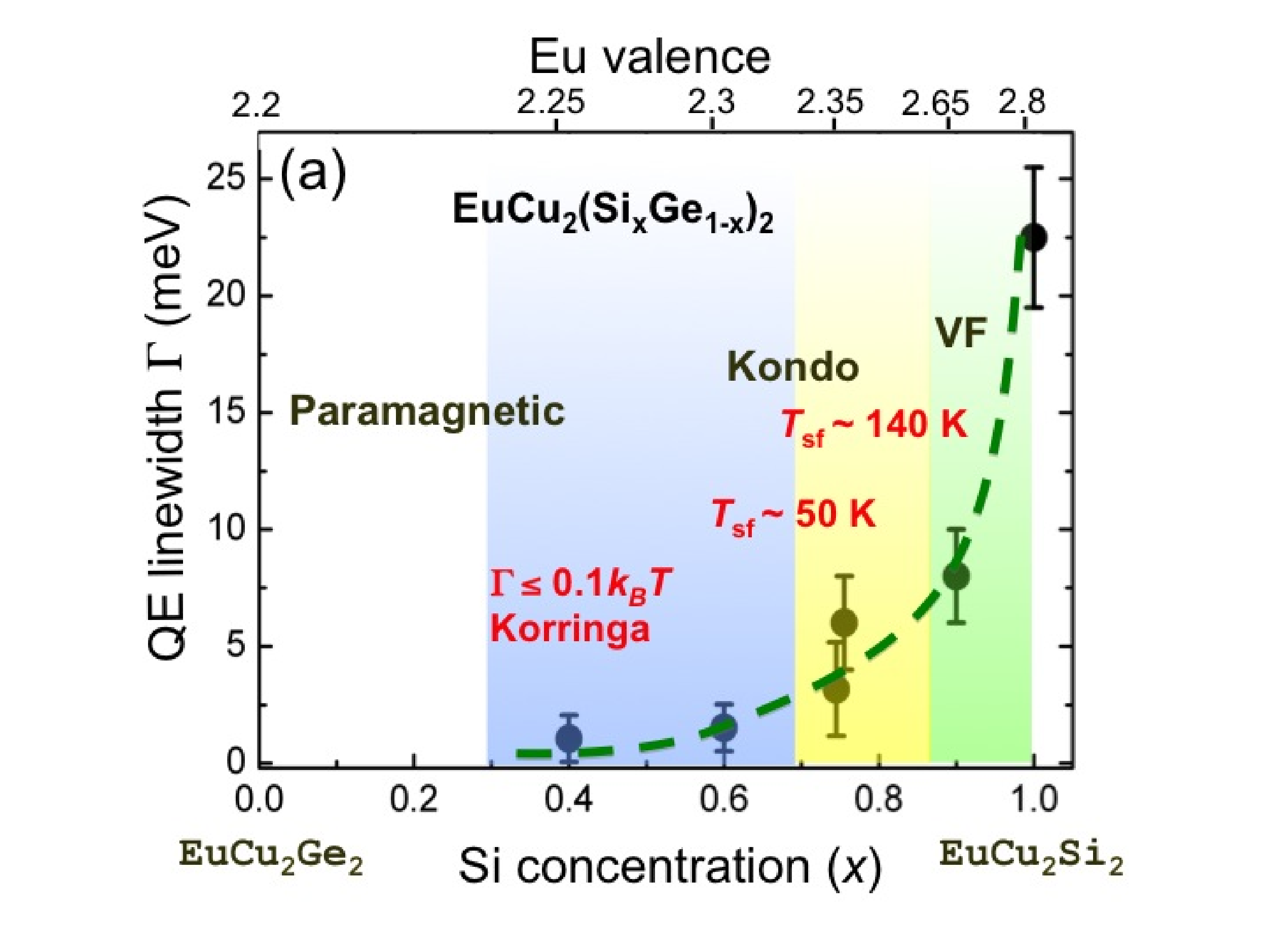}
\includegraphics[angle=0,width=0.48\textwidth]{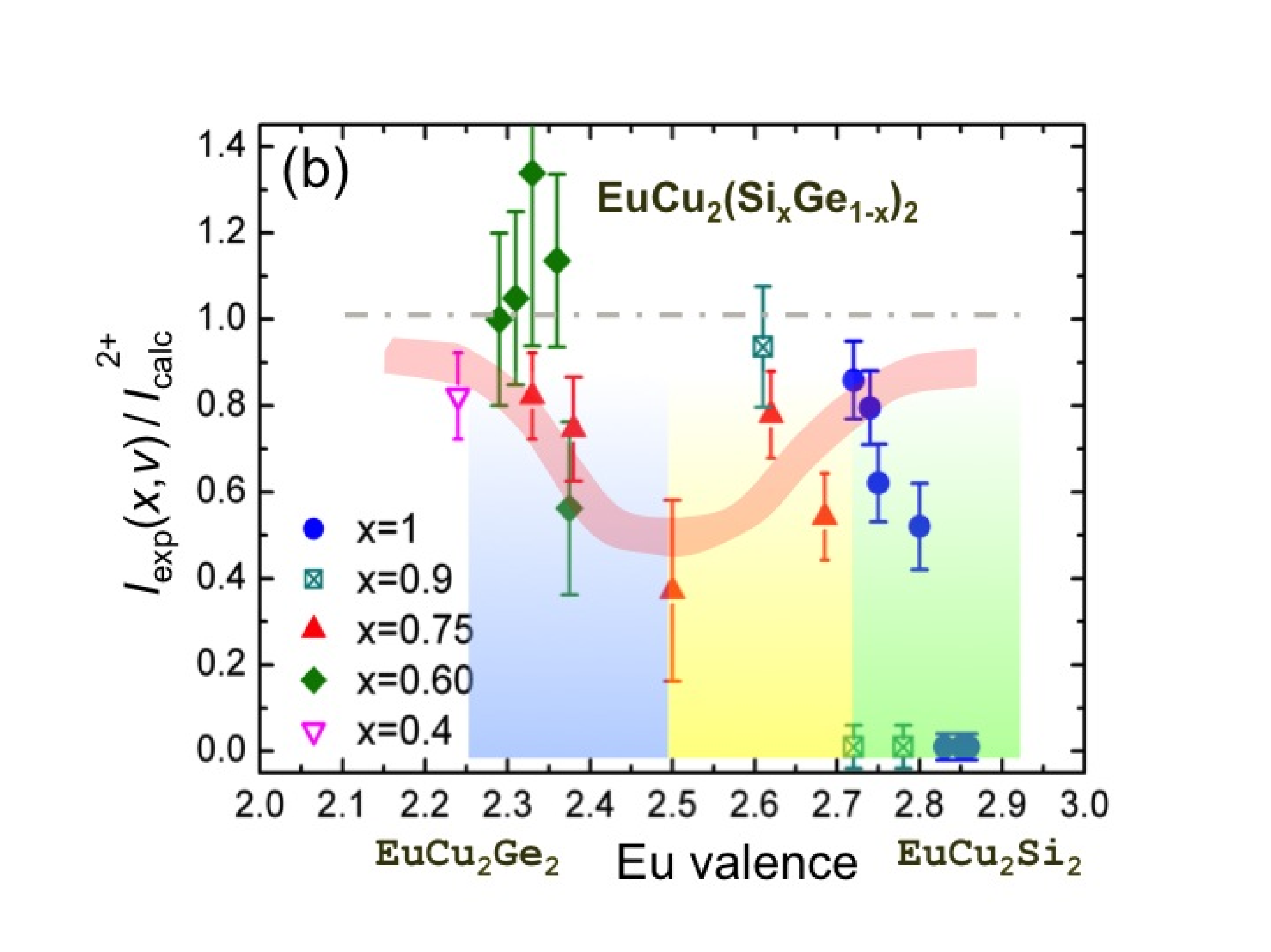}
\caption{\label{qescatt} (Color online) (a) Concentration dependence of the magnetic quasielastic linewidth (FWHM) at $T = 100$--200 K (see text) in the \ECSGx\ series. An approximate valence scale, derived from the XANES and Mössbauer results (see \Fig \ref{xan-moss} and Ref.~\onlinecite{Alekseev'12}), as discussed in Section \ref{ssec:euval} below, is indicated on the upper horizontal axis. (b) Integrated intensity of the quasielastic signal as a function of the average Eu valence derived from XANES; the values have been normalized, for each temperature and composition, to the scattering intensity expected from the estimated Eu$^{2+}$ fraction. The pink shaded trace emphasizes the general trend in the high-temperature limit (see text for details). In (a) and (b) the colored regions represent the different regimes occurring at low temperature; from right to left: (i) spin-gap, (ii) HF, (iii) long-range magnetic order (with a high-temperature paramagnetic regime characterized by normal Korringa-type thermal relaxation).}
%\vspace{-0.6cm}
\end{figure*}

The sample with the higher Si content ($x  = 0.75$) is located above the critical concentration x$_{c} = 0.65$ in the phase diagram,\cite{Hossain'04} and thus does not order magnetically. At the base temperature [\Fig \ref{ins-mag}(a)], its magnetic response contains both inelastic and quasielastic (QE) components. The former consists of two peaks, reminiscent of those observed previously in pure \ECS\ (Ref.~\onlinecite{Alekseev'07}) and EuCu$_{2}$Si$_{1.8}$Ge$_{0.2}$ (Ref.~\onlinecite{Alekseev'12}), but strongly damped, as is commonly observed in the HF regime,\cite{Knopp'89,Severing'89a,Severing'89b} to which this compound is thought to belong. The magnetic QE signal has a Lorentzian lineshape with a full width at half maximum (FWHM) $\Gamma$ of about 1.5 meV. Its existence contrasts with the spin-gap response observed\cite{Alekseev'07,Alekseev'12} below 100--150 K for compositions $x > 0.75$, and reflects a qualitative change in the energy spectrum of spin fluctuations. With increasing temperature, the QE linewidth gradually increases to exceed 3 meV at 100 K. This broadening, however, remains limited in comparison with the linewidths of 10 meV or more observed above 150 K in the Si-rich compounds \cite{Alekseev'07,Alekseev'12} [\Fig \ref{qescatt}(a)].

At lower Si concentrations ($x = 0.6 < x_{c}$), one enters the long-range magnetic order region of the phase diagram. At $T = 3$ K [\Fig \ref{ins-mag}(b)], no evidence remains for the two broad excitations previously observed above 10 meV, and the main component of the magnetic response now consists of a rather narrow QE signal ($\Gamma \approx 0.3$ meV). The asymmetry visible on the experimental spectrum is due to the detailed-balance factor. On heating, the linewidth increases to $\Gamma = 0.8$ meV at 50 K and 1.3 meV at 100 K. An extra peak near 2.4 meV, observed only in the ordered magnetic state below $T_N \approx 17$ K, likely reflects the existence of a magnon branch with a gap at the ordering wave vector $\bm{q}_{AF}$. This signal is rather weak and narrow, and merges into the magnetic excitation continuum for $T \geq T_N$. For EuCu$_{2}$Si$_{0.8}$Ge$_{1.2}$ ($x = 0.4$, not presented in \Fig \ref{ins-mag}) the spectra are quite similar to those for $x = 0.6$, apart from a further reduction of the QE linewidth at low temperature, estimated to be less than 0.25 meV.

%---------------------------------------------------------------------------------------------------------------------------------------------------------------------------------------------------

It is interesting to follow the evolution of the QE \textit{linewidth} as the Eu valence increases with increasing Si content. To avoid complications due to the spin-gap formation in the Si-rich compounds at low temperature, we focus on the QE response in the temperature region $100 \leq T \leq 200$ K. The values of $\Gamma$ derived from the present data are plotted in \Fig \ref{qescatt}(a), together with those obtained previously\cite{Alekseev'12} for \ECS\  and EuCu$_{2}$Si$_{1.8}$Ge$_{0.2}$, as a function of the Si concentration $x$. For $x = 0.4$, the linewidth is too small to be measured precisely within instrumental resolution. It remains low, of the order of 1.5 meV, in the concentration range $0.4 \leq x \leq 0.6$, then increases to 5 meV on entering the Kondo/HF regime ($x = 0.75$). Finally a dramatic rise, by a factor of five, takes place in the narrow interval $0.9 \leq x \leq 1$, leading to a value of about 22 meV in \ECS\ . This highly nonlinear dependence reflects the dependence of the average Eu valence on the Si concentration plotted in \Fig \ref{xan-moss}(c), supporting the idea that the parameter controlling the evolution of the spin dynamics across the series is the Eu valence state. At low temperature, where the QE signal exists, its width never exceeds 2--3 meV but, in the HF regime ($x = 0.75$), this relatively narrow signal coexists with quite broad inelastic peaks (linewidths of the order of 10 meV) at energies comprised between 10 and 20 meV.

The \textit{integrated intensity} of the QE signal for all measured Si concentrations and temperatures is summarized in \Fig \ref{qescatt}(b) as a function of the Eu valence. As implied by \Fig \ref{xan-moss}(c), lower valence values for a given composition correspond to higher temperatures. For the Si-rich compounds ($x = 0.9$ and 1.0), data points (shown as shaded symbols) corresponding to spectra measured in, or close to, the spin-gap regime are affected by the transfer of spectral weight from the QE to the inelastic component and therefore irrelevant to the present argument. In the spin-fluctuation regime, the general trend is tentatively represented by the pink trace. It appears that, on both ends of the plot, the measured intensity is close to that expected from the cross section calculated for the Eu$^{2+}$ fraction, whereas some reduction seems to occur in the intermediate region, where the valence strongly mixed (HF regime).

\section{\label{sec:discuss}Discussion}

\subsection{\label{ssec:euval} Eu mixed valence}
The results presented in Section~\ref{ssec:resxanes} confirm the pronounced composition and temperature variation of the Eu valence in this series of compounds, as was emphasized in previous studies.\cite{Fukuda'03,Hossain'04} However, quantitative determinations using different experimental techniques have remained controversial. Therefore, before discussing the dependence of the magnetic properties on the degree of valence mixing, one needs to consider possible problems in the interpretation of XANES and Mössbauer data.

One striking point in the XANES spectra is the existence of a sizable Eu$^{3+}$-like component for \textit{all} compositions, including pure \ECG, in which the implied deviation from divalency is far beyond the uncertainty of the method. This result, however, seems to contradict the overall ``Eu$^{2+}$-like'' behavior observed in bulk properties (entropy, magnetization). This discrepancy was already noted in the paper by Fukuda \etal,\cite{Fukuda'03} but no explanation was proposed. It has been argued, in earlier \xray\ absorption studies of other Eu intermetallics such as EuPd$_2$P$_2$ (Ref.~\onlinecite{Sampathkumaran'85}) or Eu 1-1-1 noble-metal pnictides\cite{Michels'94}, that \textit{final-state} effects can produce an artifact peak, simulating a Eu$^{3+}$ contribution, in the XANES spectra of purely divalent compounds. In such a process, one electron from the $4f^7$ shell is partially promoted (``shake-up'') into one of the ligand orbitals, following the creation of a $2p$ core hole by the incoming photon. This is more likely to occur in systems with a higher degree of covalency, as may be the case close to a valence instability, where $4f$ states hybridize with ligand orbitals.\cite{Wortmann'86} However, there is no consensus so far on the possible magnitude of such effects in one given material.  On the other hand, it is known that some divalent 1-2-2 Eu compounds (EuFe$_2$As$_2$, Refs.~\onlinecite{Sun'10,Matsubayashi'11}, or EuCo$_2$As$_2$, Ref.~\onlinecite{Tan'16}) exhibit a single Eu$^{2+}$ peak under normal conditions, while they develop a two-peak XANES structure when a MV state is produced by means of hydrostatic or chemical pressure. The role of final-state effects in that class of systems has been questioned by Röhler\cite{[{}] [{, in particular see p. 521 sqq.}]{Rohler'87}} and remains partly unsettled. 

Quantitatively, one can note that all deviations from divalency ascribed to final-state effects in Ref.~\onlinecite{Sampathkumaran'85} (fractional Eu$^{3+}$ intensity of approximately 15\% in EuPd$_2$P$_2$), Ref.~\onlinecite{Michels'94} (apparent valence comprised between 2.12 and 2.22 in 1-1-1 compounds), and Ref.~\onlinecite{Wortmann'86} (deviation of 0.09 to 0.12 in the EuPd$_{2-x}$Au$_x$Si$_2$ series) are weak in comparison with those observed in the \ECSGx\ compounds, especially for $x \geq 0.4$. Furthermore, for $x = 0.75$ and, notably, for $x = 0.6 < x_c$, the temperature dependence of the valence determined from XANES (also observed by Fukuda \etal \cite{Fukuda'03} for the same composition, and even faintly for $x = 0.5$) supports a true valence-mixing effect.

In Section~\ref{ssec:resxanes}, the Mössbauer results were used primarily to demonstrate the homogeneous character of the MV state. Valence determination based on the IS, on the other hand, is problematic. The main problem comes from the lack of reliable di- or trivalent reference system. In the present work, for x $\leq$ 0.75 we found IS comprised between $-9.8$ and $-7.5$ mm/s, with IS = $-8.6$ mm/s for $x = 0.6$. Michels \etal \cite{Michels'94}, have reported isomer shifts near $-10.7$ mm/s or below for divalent EuAuP and EuCuPt, although they mention relative velocities covering a wide range between $-12$ and $-8$ mm/s for Eu$^{2+}$ in other metallic compounds. The values found here are systematically larger than the average estimate for divalent Eu, though still in a range compatible with a pure Eu$^{2+}$ state.

Based on the literature data for the compressibility of EuCu$_{2}$Si$_{2}$,\cite{Neumann'85} the expansion of the lattice due to the substitution of Ge for Si,\cite{Hossain'04} and the typical pressure dependence of the Eu isomer shift of ~10$^{-2}$mm$\cdot$s$^{-1}$$\cdot$kbar$^{-1}$ (Ref.~\onlinecite{Grandjean'89}), we have estimated the possible change in the Eu isomer shift due to the difference in lattice parameters within the EuCu$_{2}$(Si$_{x}$Ge$_{1-x}$)$_{2}$ series. We have found that the expected change in the isomer shift for EuCu$_{2}$Si$_{0.8}$Ge$_{1.2}$ and EuCu$_{2}$Si$_{1.2}$Ge$_{0.8}$ with respect to EuCu$_{2}$Ge$_{2}$ is about 0.35 and 0.45 mm/s, respectively, whereas the values obtained experimentally are 0.7 and 1.2 mm/s, respectively, are at least twice higher. Therefore even if EuCu$_{2}$Ge$_{2}$ is assumed to be divalent, this suggests that other compositions are mixed-valent. 
Although this difference cannot be regarded as a conclusive proof, it lends support to our assumptions, at least for $x \geq 0.6$.

In summary, we believe that the present XANES and IS data consistently point to the existence of a MV state of Eu in \ECSGx\ for $x = 0.6$ and above, in particular in the region of interest, near $x = x_c$, where competing Kondo, HF, and long-range-order phenomena have been reported to occur.\cite{Hossain'04} The large trivalent contribution in the XANES spectra, as well as its significant temperature dependence, are unlikely to result from shake-up effects. This regime probably extends to the Ge-rich range up to $1-x = 0.6$. For even higher Ge contents, as well as in pure \ECG, it is difficult to decide whether the residual (but significant) trivalent character indicated by the present, as well as Fukuda's\cite{Fukuda'03}\ earlier XANES results, is entirely due to experimental artifacts. This point is not critical to our discussion of magnetic properties, and remains open for future studies. In the following, we will not attempt to correct the valence values obtained in Section~\ref{ssec:resxanes} and use the correspondence between valence and composition as displayed in \Fig \ref{xan-moss}.

\subsection{\label{ssec:dynmag} Dynamic magnetic response}
Starting from pure \ECG, the substitution of Si causes a reduction of the ordered Eu magnetic moment, as shown by the neutron diffraction results. This moment reduction can be ascribed to the approach of the strong spin fluctuation regime ($x > 0.6$), which gradually suppresses Eu long-range magnetism and correlates with the increase in the average Eu valence evidenced from XANES experiments.

Key features of the spin fluctuations dynamics developing in the MV state are revealed by neutron spectroscopy. In a wide composition range ($0.4 < x < 0.75$), a pronounced QE peak is observed in the magnetic spectral function, in contrast with the spin-gap behavior ($\Delta \approx 20$--30 meV) developing at low temperature in pure \ECS\  ($x = 1$) and EuCu$_{2}$Si$_{1.8}$Ge$_{0.2}$ ($x = 0.9$). In Refs.~\onlinecite{Alekseev'12,Alekseev'07}, the latter compounds were shown to exhibit an inelastic response at $T = 5$ K consisting of two excitations, which were ascribed to a renormalized Eu$^{3+}$ spin-orbit excitation $^7F_0 \rightarrow {}^7F_1$ and a resonance-like magnetic mode, respectively.

Above the temperature of the spin-gap suppression (on the order of 100 K), a very high spin-fluctuation rate was observed, as already noted above. With increasing Ge content ($x$ decreasing from 1 to 0.75), the inelastic signal broadens and shifts to lower energies. In the region of maximum valence mixing (average Eu valence $v \sim  2.5$ at $T = 10$ K near $x = 0.65$), a QE signal coexists, at low temperature, with overdamped inelastic peaks [see \Fig \ref{ins-mag}(a) for $x = 0.75$]. For $x = 0.6$ and below, only the QE response exists (apart from a magnon-like component below \TN\ seen in \Fig \ref{ins-mag}(b) for $x = 0.6$). We stress that no (Eu$^{3+}$)-type inelastic peak was observed here in the spectra for $x = 0.6$ (near 45 meV), and $x = 0.4$ (up to 30 meV).

From these results and the discussion given in the previous Section, one is led to the important conclusion that the long-range order developing below $x_c$, e.g. for $x = 0.6$ in EuCu$_{2}$Si$_{1.2}$Ge$_{0.8}$, cannot be based on the magnetism of the Eu$^{2+}$ ionic component alone, but represents a genuine property of the MV state, whose character gradually changes, with decreasing $x$, from a nonmagnetic singlet to a degenerate spin-fluctuation state. In view of gradual evolution observed, as a function of composition, in the AFM region, the same possibly applies to lower Si concentrations as well. Obviously, the MV state does not preclude the occurrence of long-range order, and might even provide additional coupling channels whereby this order can develop.

\begin{figure}[t] %------------------------------------------------------------- Figure phase diagram --------------------------------------------------------------------------------
%\vspace{-0.6cm}
%\centering
\includegraphics[angle=0,width=0.8\columnwidth]{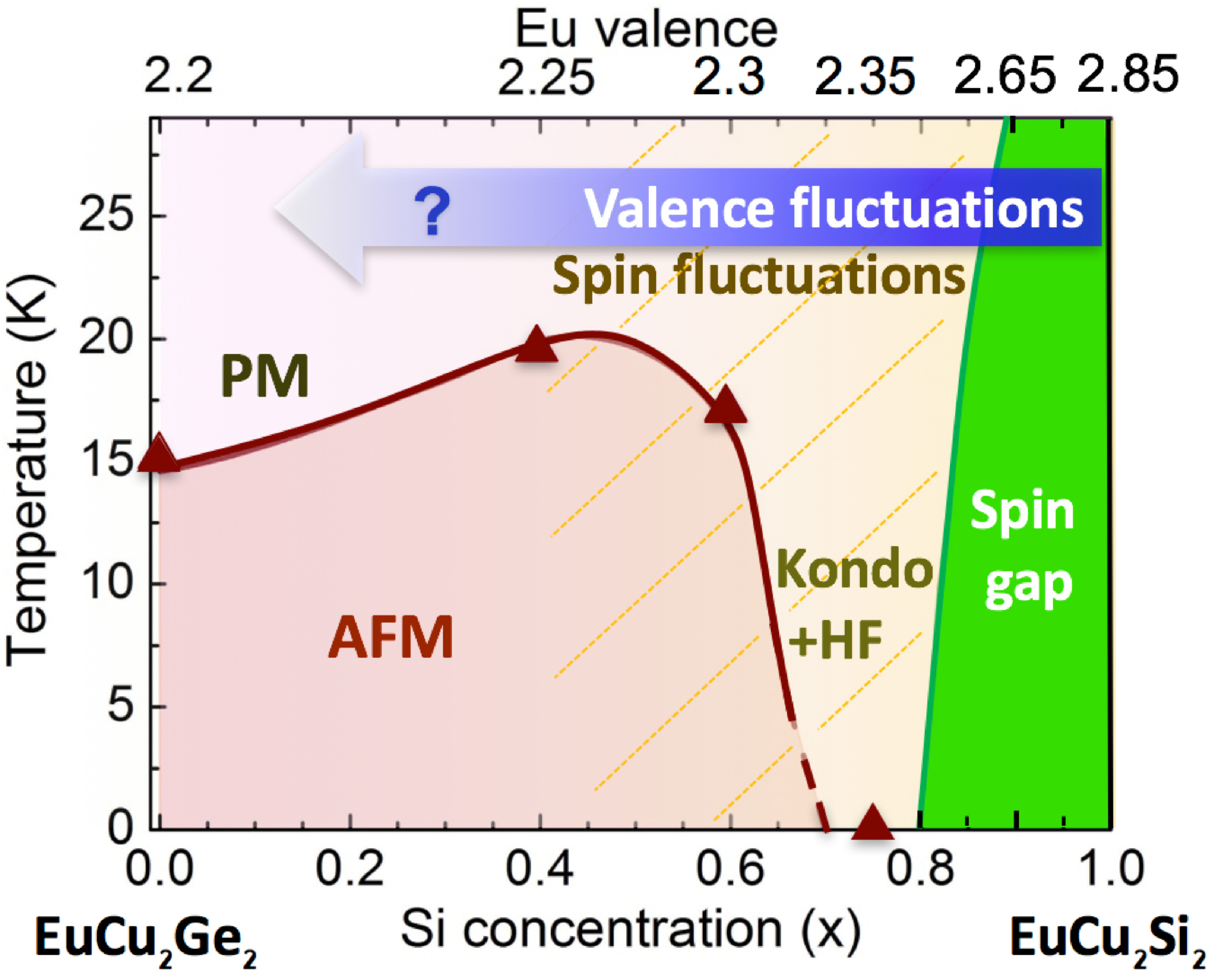}
\caption{\label{phdiag} (Color online) Magnetic phase diagram for \ECSGx\ based on the original data of Ref.~\onlinecite{Hossain'04} (solid brown line showing the phase boundary between the AFM and paramagnetic states) and those obtained in the present work. Triangles indicate values obtained from our NPD experiments and correspond to samples composition used in the present work. The yellow-colored area, extending to $x = 0.4$ and encroaching upon the AFM region, corresponds to the spin-fluctuation regime in which a QE response is observed, and the green-colored area to the spin-gap regime. The valence values indicated on the upper scale are those derived from the XANES data at $T = 7$ K of the present work and of Ref.~\onlinecite{Alekseev'12}.}
%\vspace{-0.6cm}
\end{figure}
%---------------------------------------------------------------------------------------------------------------------------------------------------------------------------------------------------

The results are mapped out on the magnetic phase diagram for the entire \ECSGx\ series (\Fig \ref{phdiag}). The Néel temperatures obtained from neutron diffraction, represented by triangles, agree perfectly with those reported in the previous work,\cite{Hossain'04} and the observation of magnetic superstructure peaks further confirms the long-range character of the order. In particular, we note the unexpected robustness of the AFM ordered state, characterized by an initial increase in the Néel temperature from 15 to 19 K between $x = 0$ and $\sim$\,0.4, followed by a moderate decrease to 17 K at $x = 0.6$, despite the reduction of the Eu magnetic moment likely due to the enhancement of spin fluctuations [\Fig \ref{qescatt}(a)]. The order observed at the composition $x = 0.6$ is of particular interest because it corresponds to a regime in which a Kondo behavior has been clearly established\cite{Hossain'04}, with a Kondo temperature ($\TK \sim 10$ K) comparable to the Néel temperature. The absence of long-range order in the IN4C data for $x = 0.75$ agrees with the value of $x_{c} = 0.65$ reported in Ref.~\onlinecite{Hossain'04}, and confirms that the drop of \TN\ to zero occurs precipitously in a narrow concentration interval, just above $x =0.6$.

The green-colored area shown in \Fig \ref{phdiag} for Si concentrations above $x \sim 0.8$ corresponds to the formation of the spin-gap in the excitation spectra at low temperature. The yellow-colored area denotes the existence of a detectable QE magnetic signal in the INS spectra. The gradual evolution of the dynamic response as $x$ decreases provides further insight into the formation of the spin-fluctuations state near x$_{c}$. Above $x = 0.8$, the low-temperature behavior is dominated by the suppression of the low-energy spectral weight in the spin-gap range ($E < 30$ meV) below $\sim$\,150 K, associated with the formation of a Eu singlet ground state.\cite{Alekseev'12} As $x$ decreases, the inelastic components renormalize to low energies, as shown in Ref.~\onlinecite{Alekseev'12}, eventually leading to a pure QE response at $T = 50$ K $> \TN$ for $x < 0.6$. The gradual appearance of the QE signal around $x = 0.6$--0.75 reflects the recovery of magnetic moments in the ground state (instead of the singlet ground state with a spin gap for $x \geq 0.8$). This can be viewed as the necessary condition for both the enhancement of spin fluctuations and the tendency to form the long-range ordered magnetic state. Around $x = 0.75$, the decrease (in comparison with higher $x$) of the spin-fluctuation energy to the range of a few millielectronvolts allows the HF state to be formed, leading to the first known case of a HF behavior occurring in a strongly MV material (around $v \sim 2.5$--2.6, according to XANES).

The competition between, and/or coexistence of, long-range magnetic order and strong spin fluctuations has been extensively studied in Ce or Yb-based intermetallic compounds, a number of which\cite{Knopp'89,Severing'89a,Severing'89b} belong to the same 1-2-2 family as \ECSGx. In such compounds, the lanthanide ion occurs in a nearly trivalent HF state, with a magnetic (degenerate) ground state defined by the crystal field splitting, and the properties are well understood in terms of the competition between the Kondo effect and RKKY exchange interactions, as suggested in Doniach's\cite{Doniach'77} and subsequent models.  Strongly MV Ce compounds, on the other hand, do not exhibit magnetic order. The latter scenario is in strong contrast with the present situation, where the AFM order extends far into the MV regime, and its suppression occurs close to one-to-one mixing of the Eu$^{2+}$ and Eu$^{3+}$ states.

A clue to clarifying the similarities and differences between ``classical'' Ce$T_{2}$$X_{2}$ systems and MV \ECSGx\ may be given by the magnetic spectral response observed in the present INS study. It is important to note that, both in the case of Ce and Yb, one of the electronic configurations involved in the valence fluctuation is nonmagnetic not just as a result of Russel-Saunders and/or spin-orbit coupling, but because its 4$f$ shell is either empty (Ce) or full (Yb). In Eu$^{3+}$, on the other hand, the ionic ground-state multiplet is indeed a singlet ($^7F_0$) but, as revealed by the INS spectra, magnetic (spin-orbit) excited states exist at relatively low energies, less than 40 meV in pure \ECS. As Ge is substituted for Si, this energy further decreases, while spectral weight is gradually transferred to the QE region. Meanwhile, the spin fluctuation rate, evidenced by the linewidth of the QE signal at high temperature, decreases considerably. The emergence of a degenerate ground state due to the renormalization to low energies of the inelastic part of the Eu spectrum, as well as the slowing down of spin fluctuations, may restore the conditions for Kondo-type ($s$-$f$ exchange) spin dynamics in competition with long-range magnetic order. This may also explain why, contrary to the Ce case, the strongly MV character does not preclude the emergence of the magnetic state.

That magnetic order can develop in the presence of strong valence fluctuations, provided magnetic degrees of freedom exist in both valence states, was well documented, back in the 1980s, for the rock-salt structure chalcogenide compound TmSe. Despite the Tm valence being strongly noninteger ($v \sim 2.58$, almost temperature independent), type-I AFM order was found to set in below $\TN = 3.45$ K for stoichiometric samples\cite{Bucher'75,BjerrumM'77,Launois'80}. Thulium shares with europium the multiple electron occupancy of its $4f$ shell, unlike cerium and ytterbium, which have only one $4f$ electron (Ce) or hole (Yb) in their trivalent ionic states. On the other hand, the peculiar MV behavior of TmSe is generally ascribed to the fact that the ground state multiplets of both Tm$^{2+}$ ($4f^{13}$, $^2F_{7/2}$, $p_{\mathrm{eff}} = 4.5\mub$) and Tm$^{3+}$ ($4f^{12}$, $^3H_{6}$, $p_{\mathrm{eff}} = 7.5\mub$) configurations are magnetic (neglecting crystal-field effects\cite{Furrer'81}, which are likely wiped out by the valence fluctuations). This ingredient is central to several of the models\cite{Loewenhaupt'79,Mazzaferro'81,Alascio'82,Schlottmann'81,HollandMoritz'83} proposed to explain the properties of TmSe and, obviously, cannot be carried over to \ECSGx\ where, as already noted, valence mixing involves one magnetic Eu$^{2+}$ ($J = 7/2$) and one non-magnetic Eu$^{3+}$ ($J = 0$) ionic configurations.

The magnetic spectral response of TmSe\cite{Loewenhaupt'79,HollandMoritz'83} also differs significantly from that observed in \ECSGx\ below the critical concentration. Above $T = 100$ K, a rather broad ($\Gamma \sim 10$ meV,\cite{Loewenhaupt'79,HollandMoritz'83} comparable to $\Gamma = 22$ meV in \ECS\  in the same temperature range), temperature-independent QE response is observed, reflecting the existence of fast spin fluctuations but, upon cooling, the QE linewidth decreases with a crossover to a linear temperature dependence, $\Gamma / 2 \sim 0.7 k_BT$, for $T \rightarrow 0$. Simultaneously an inelastic response appears, whose energy increases on cooling to reach about 10 meV at $T = 10$ K. The intensity of the latter mode exhibits a strong periodic $Q$ dependence, with a maximum at the fcc zone-boundary $X$ point. Below \TN, the QE scattering is suppressed.

In 1-2-2 intermetallics, electron states of different symmetries ($s$, $d$) can occur at the Fermi level\cite{Sticht'86,Jarlborg'83}. In TmSe, on the other hand, the only electrons populating the conduction band are those provided by the hybridization with the $4f$ orbitals. As a result, TmSe is known to exhibit unique Kondo-insulator properties at low temperature,\cite{Haen'79} whereas the present compounds remain metallic. Accordingly, the suppression of the QE signal due to spin fluctuations in the AFM state below \TN\cite{Loewenhaupt'79} is at variance with the behavior observed here in \ECSGx, as well as in Ce-based HF compounds.

Theoretical attempts to specifically address the multiple occupancy of the $4f$ shell in MV systems are rather scarce. Apart from those applicable to TmSe, which rely on the existence of two magnetic valence states, and are therefore not directly relevant to the case of Eu, one can mention the work of Bulk and Nolting\cite{Bulk'88}, developed in connection with early experimental results on Eu systems (elemental Eu, Eu[Pd,Au]$_2$Si$_2$). Their extended ``\hbox{$s$--$f$} model'' considers, in addition to the hybridization, $V$, between the $4f$ and the conduction band states, an independent, non-Kondo, \hbox{$s$--$f$} exchange interaction $J_{sf}$ assumed to be positive (i.e. ferromagnetic) in the case of Eu compounds. The AFM order is then ascribed to a direct exchange coupling $J_{1}^{AB}$ between $4f$ magnetic moment at neighboring Eu sites. This model accounts for the possibility of developing AFM order inside the MV regime. It also predicts, for some parameter range, that the Néel temperature can increase with increasing $V$, as observed in the \hbox{low-$x$} region of the \ECSGx\ phase diagram. However, the clear observation of a Kondo-type behavior in the electrical transport coefficients,\cite{Hossain'04} seems to rule out the predominantly FM $s$-$f$ coupling assumed in that model. 

In a recent paper,\cite{Hotta'15} Hotta has proposed an interesting theoretical basis to explain how the $4f^7$ state of Eu$^{2+}$ can give rise to a Kondo phenomenology (including quantum criticality controlled by an external parameter) very similar to that found in nearly trivalent ($4f^1$) Ce compounds. The key argument is that, for realistic values of the spin-orbit coupling, a correct description of atomic $4f$ states cannot be achieved in terms of the standard Russel-Saunders scheme. The real situation is intermediate between $LS$ and $j$--$j$ couplings and, even for a relatively weak spin-orbit interaction, $\lambda_{\mathrm{so}}/U \sim 0.1$ ($U$: Hunds rule interaction), i.e. far from the pure $j$--$j$ regime, this has to be taken into account. The main result reported for this regime is the observation of a ``single-$f$-electron''-like behavior, due to 6 electrons being accommodated in a fully occupied $j = 5/2$ sextet ($j = l - s$),  while one single electron occupies the $j = 7/2$ octet ($j = l + s$). The latter state can account for a $R \mathrm{ln}2$ step in the entropy, a (Yb-like) $\Gamma_6$--$\Gamma_7$--$\Gamma_8$ crystal-field scheme, as well as for Kondo effect. The model has been applied\cite{Hiranaka'13,Mitsuda'13} to the HF behavior reported in EuNi$_2$P$_2$. The \ECSGx\ series could provide a second useful benchmark for these ideas. To this end, Hotta's approach should be extended to take proper account of the degenerate conduction bands (multi-channel Kondo?), and the consideration of valence fluctuations affecting the Kondo regime. It has also been suggested that the treatment of MV in Eu should consider ``inter-site'' Coulomb repulsion (à la Falicov-Kimball\cite{Zlatic'03}) between local and conduction electrons, which is not included in the Anderson model.

The evolution of the spin dynamics observed in the present study on approaching the critical concentration provides guidelines along which further theoretical work should be undertaken. A possible starting point is the previous description \cite{Alekseev'12} of the magnetic response for pure \ECS\  and the Si-rich solid solutions in terms of a renormalized spin-orbit excitation associated with the parent Eu$^{3+}$ configuration, with an extra magnetic exciton mode below the spin-gap edge, as proposed in Ref.~\onlinecite{Kikoin'95}.

\section{\label{sec:conclusion}Conclusion}

In summary, the \ECSGx\ MV system exhibits an unusual ground state involving a coexistence of long-range antiferromagnetic order and spin fluctuations, observed over a significant concentration range. The critical value $x_c = 0.65$ corresponds to the suppression of the magnetic order and the appearance of a HF behavior. This observation is at variance with the typical behavior found in Ce- and Yb- based 1-2-2 HF systems, and requires further theoretical understanding. The analysis of the inelastic and QE magnetic contributions to the Eu magnetic spectral function provides clues as to the physical mechanism of the crossover from spin fluctuations to magnetic order, and the origin of the HF state in this unconventional situation. In particular, we emphasize the evolution of the magnetic response of MV Eu, as the Ge content increases, from the spin-gap spectrum found in pure \ECS\  to a degenerate ground state with moderate spin fluctuations. This evolution takes place through a renormalization of the magnetic excitations to lower energies and the transfer of spectral weight to the quasielastic component. This spectral rearrangement favors the formation of a HF ground state in the corresponding intermediate region of the phase diagram. Below \TN, spin fluctuations extend into the long-range order state. An important open question, in analogy with the Ce- and Yb-based 1-2-2 HF systems, is the possible existence of quantum criticality near the AFM onset. Addressing this question by means of neutron scattering would require a detailed single-crystals study of the $\bm{Q}$ dependence of the fluctuations.

%\section{\label{sec:ack}Acknowledgements}
\begin{acknowledgments}
We are grateful to C. Geibel, I. P. Sadikov, V. N. Lazukov, A. V. Kuznetsov, I. Sergueev and R. Gainov for hepful discussions, to B. Beuneu, C. Pantalei and A. Orecchini for their support during the neutron scattering experiments, to J. de Groot and H. Nair for the help in the magnetic sample quality assessment, to I. Zizak and D. Wallacher for assistance in the XANES experiment at BESSY II, to DESY Photon Science and Helmholtz Zentrum Berlin for providing beamtime. Support from the Helmholtz Gemeinschaft Deutscher Forschungszentren for funding of the Helmholtz-University Young Investigator Group NG-407 (RPH), and from the U.S. Department of Energy, Office of Science, Basic Energy Sciences, Materials Sciences and Engineering Division (RPH) is acknowledged. The work was partly supported by the grants of RFBR No. 14-22-01002 and 14-02-01096-a (neutron measurements), and Russian Science Foundation No. 14-22-00098 (synchrotron radiation measurements).
\end{acknowledgments}

% --------- Insert bibliography from external BibTeX library 
%\bibliography{Nemkovski_EuCuSiGe_PRB} 

%% --------- In-text bibliography 
% >>> COMPLETE REFERENCE WORTMANN'86 BY ADDING "-879" BY HAND TO PAGE NUMBER "C8" <<<
%

\end{document}